\documentclass[10pt, conference, compsocconf]{IEEEtran}
\usepackage[utf8]{inputenc}
\ifCLASSINFOpdf
    \usepackage[pdftex]{graphicx}
    \graphicspath{{./images/}}
    \DeclareGraphicsExtensions{.pdf}
\else
\fi
\usepackage[cmex10]{amsmath}
\usepackage{array}
\usepackage[caption=false,font=footnotesize]{subfig}
\usepackage{amssymb}
\usepackage{todonotes}
\usepackage{acronym}

\acrodef{IP}{Internet Protocol}
\acrodef{DNS}{Domain Name System}
\acrodef{TLS}{Transport Layer Security}
\acrodef{CPRNG}{Cryptographic Pseudo-Random Number Generator}
\acrodef{GPS}{Global Positioning System}
\acrodef{geocast}{Geographic Multicast}
\acrodef{mobicast}{Just-in-time Multicast}
\acrodef{STM}{Spatiotemporal Multicast}
\acrodef{GTM}{Geo-temporal Multicast}
\acrodef{CSTM}{Cluster-based Spatiotemporal Multicast}
\acrodef{SMS}{Short Message Service}
\acrodef{LTE}{Long-Term Evolution}
\acrodef{S}{Sender}
\acrodef{UE}{User Equipment}
\acrodefplural{UE}[UEs]{User Equipments}
\acrodef{eNB}{evolved NodeB}
\acrodef{HSS}{Home Subscriber Server}
\acrodef{MME}{Mobility Management Entity}
\acrodef{UE}{User Equipment}
\acrodef{PGW}{Packet Data Network Gateway}
\acrodef{IMSI}{International Mobile Subscriber Identity}
\acrodef{TMSI}{Temporary Mobile Subscriber Identity}
\acrodef{NAT}{Network Address Translation}
\acrodef{RP}{Rendezvous Point}
\acrodefplural{RPs}[RPs]{Rendezvous Points}
\acrodef{DoS}{Denial-of-Service}
\acrodef{TPS}{Token Planning Server}
\acrodef{st-r}[$st$-region]{spatiotemporal region}
\acrodef{st-d}[$st$-datagram]{spatiotemporal datagram}
\acrodef{MSN}{Mobile Social Networking}
\acrodef{GDH}{Group Diffie-Hellman}
\acrodef{ECDH}{Elliptic Curve Diffie-Hellman}
\acrodef{VUE}{Voting for Urgent Events}
\acrodef{DTN}{Delay-Tolerant Network}
\acrodef{HMAC}{Hash-based Message Authentication Code}
\acrodef{IS}{Identity Server}
\acrodef{RWP}{Random Waypoint}
\acrodef{RPGM}{Reference Point Group Mobility}
\acrodef{NC}{Nomadic Community}
\acrodef{DTN}{Delay-Tolerant Network}

\newcommand{\itembf}[1]{\item\textbf{#1:} }

\usepackage[absolute]{textpos}
\usepackage{calc}
\newcommand{\copyrightnotice}{
    \begin{textblock}{0.78}(0.11,0.03)
        \noindent
        \footnotesize
        \textcolor{gray}{
        {\bf This paper is a preprint (IEEE "accepted" status).} 
        ~\copyright~2013 IEEE. 
        Personal use of this material is permitted.
        Permission from IEEE must be obtained for all other uses, in any current or future media, including reprinting/republishing this material for advertising or promotional purposes, creating new collective works, for resale or redistribution to servers or lists, or reuse of any copyrighted component of this work in other works.
        }
    \end{textblock}
}

\begin{document}
\title{Towards Trustworthy Mobile Social Networking Services for Disaster Response}
%------------------------------------------------------------------------------


% \author{\IEEEauthorblockN{Authors Name/s per 1st Affiliation (Author)}
% \IEEEauthorblockA{line 1 (of Affiliation): dept. name of organization\\
% line 2: name of organization, acronyms acceptable\\
% line 3: City, Country\\
% line 4: Email: name@xyz.com}
% \and
% \IEEEauthorblockN{Authors Name/s per 2nd Affiliation (Author)}
% \IEEEauthorblockA{line 1 (of Affiliation): dept. name of organization\\
% line 2: name of organization, acronyms acceptable\\
% line 3: City, Country\\
% line 4: Email: name@xyz.com}
% }

%------------------------------------------------------------------------------
\author{
\IEEEauthorblockN{Sander Wozniak, Michael Rossberg, Guenter Schaefer}
\IEEEauthorblockA{Ilmenau University of Technology\\
\{sander.wozniak, michael.rossberg, guenter.schaefer\}[at]tu-ilmenau.de}
}
%------------------------------------------------------------------------------


%Telephone: (800) 555--1212, Fax: (888) 555--1212}


\maketitle

% add copyright notice to preprint
\copyrightnotice

%------------------------------------------------------------------------------
\begin{abstract}
Situational awareness is crucial for effective disaster management.
However, obtaining information about the actual situation is usually difficult and time-consuming.
While there has been some effort in terms of incorporating the affected population as a source of information, the issue of obtaining trustworthy information has not yet received much attention.
Therefore, we introduce the concept of witness-based report verification, which enables users from the affected population to evaluate reports issued by other users.
We present an extensive overview of the objectives to be fulfilled by such a scheme and provide a first approach considering security and privacy.
Finally, we evaluate the performance of our approach in a simulation study.
Our results highlight synergetic effects of group mobility patterns that are likely in disaster situations.
\end{abstract}
%------------------------------------------------------------------------------

\begin{IEEEkeywords}
Security and Privacy Protection, Mobile communication systems, Multicast
\end{IEEEkeywords}

% \fi
\IEEEpeerreviewmaketitle

%------------------------------------------------------------------------------
\section{Introduction}
\label{sec:intro}

Responding to large-scale disasters has always been a challenging task.
One of the reasons for this is the unpredictability of the actual situation at hand.
With first responders usually being short on technical and human resources, an awareness of the current circumstances, e.g. the location of casualties, is substantial to effectively providing help to victims within the first critical hours.
In order to increase the situational awareness of officials and to support mutual first response, the concept of incorporating the affected population as a potential source of information has emerged recently \cite{palen2010vision}.
Among the potential \ac{MSN} services for disaster response \cite{wozniak2011towards}, one of the most important services is a reporting service that enables the affected population to issue reports about the locations of victims, remaining or evolving hazards, resource requirements, etc.
With other services building upon the data collected by this service, it is essential that this information is authentic and accurate to allow appropriate decision making.
Therefore, apart from ensuring a high quality of information, a crucial aspect of this service is to implement countermeasures against users trying to inject false or inaccurate information about allegedly urgent events.

\begin{figure}[!t]
\centering
\includegraphics{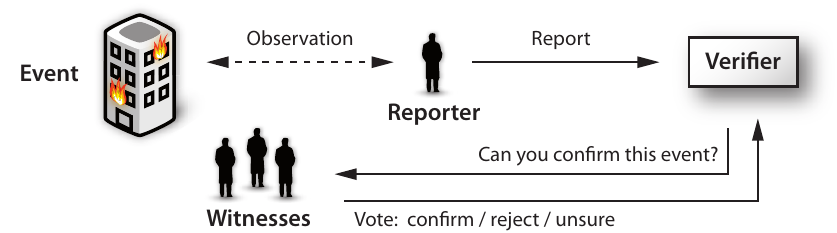}
\caption{Witness-based report verification}
\label{fig:concept}
\end{figure}

In this work, we introduce a rating approach relying on the affected population to verify the correctness and urgency of reports.
In our approach, which we refer to as \ac{VUE}, witnesses report certain events to so-called verifier nodes.
These verifier nodes issue confirmation requests to potential witnesses of the event, asking them to decide about the accuracy and urgency of the report.
Witnesses can then vote with their decision, allowing the verifier node to rate a report (see Fig.\;\ref{fig:concept}).

Our witness-based approach is inspired by the issue of obtaining credible information in \emph{social swarming} applications \cite{liu2011optimizing}.
In social swarming, a swarm of users tries to cooperatively fulfill certain tasks, e.g., search and rescue.
Users in the swarm may send reports to a swarm director using their smartphones.
Based on his global view, the swarm director then provides instructions to users to achieve the common goal.
In order to obtain credible information, the swarm director may selectively query users for confirmation.
Accordingly, in our verification schemes, confirmation requests are issued to certain users.
However, in their work, the authors focus on the problem of optimizing the network resources by querying the most suitable users based on their credibility under normal network conditions.
In contrast, we apply the concept of querying specific users to deal with the challenges of verifying reports in disaster situations.
On one hand, this concerns the need to communicate in a delay-tolerant manner due to the failure of parts of the network infrastructure.
On the other hand, in order to meet legal requirements and gain acceptance among users, such a scheme has to protect the privacy of the witnesses.
This is especially the case if such an approach is deployed on mobile devices that are also used in normal conditions, e.g., to provide help also in a small scale car accident.

Apart from the issue of obtaining credible information in social swarming, there are several related research areas.
On one hand, there has been work on trustworthy ubiquitous emergency communication \cite{weber2011mundomessage}.
However, it focuses on first responders and does not consider the verification of information for \ac{MSN} services.
On the other hand, regarding the issue of crowdsourcing information in disasters, existing approaches are usually open-access, with no or only limited verification \cite{gao2011harnessing}.
Furthermore, while there has been work on the trustworthiness of information obtained from microblogging services for emergency situations \cite{gupta2012credibility}, the aspect of querying witnesses in the disaster area in order to verify reports has not been considered yet.
Finally, our approach can be considered an application of the concept of \emph{spatiotemporal multicast}, where a message is delivered to users, i.e., witnesses, encountered in the past while protecting their privacy from the sender of the message \cite{wozniak2012geocast}.

In this article we make the following contributions:
We propose the concept of witness-based report verification in the context of a reporting service for disasters and derive extensive security and privacy objectives (section\;\ref{sec:objectives}).
Furthermore, we present a first approach for such a scheme (section\;\ref{sec:approach}) and provide a detailed discussion of its security and privacy features (section\;\ref{sec:discussion}).
Finally, we evaluate our approach by an extensive simulation study (section\;\ref{sec:evaluation}).

%------------------------------------------------------------------------------
\section{Design Objectives}
\label{sec:objectives}
%------------------------------------------------------------------------------

In this work, we consider a network model where users are able to sporadically access the Internet via a cellular network infrastructure.
Furthermore, we assume that devices are able to communicate directly forming a local wireless network.

%------------------------------------------------------------------------------
\subsection{Functional Objectives}
%------------------------------------------------------------------------------

\begin{itemize}
    \itembf{Proximity restriction} Only users close to an event should be able vote for reports about this event.
    \itembf{Deferring of votes} Users should be able to defer a vote, e.g., if a user has to provide first aid, he should be able to defer his vote and submit it later.
\end{itemize}

%------------------------------------------------------------------------------
\subsection{Non-functional Objectives}
%------------------------------------------------------------------------------

\begin{itemize}
    \itembf{Verification delay} Reports should be verified quickly.
    \itembf{Robustness} After a disaster, parts of the infrastructure may fail.
        Hence, the scheme has to operate in a delay- and disruption-tolerant manner.
        Furthermore, it should be robust against occasional false reports and votes.
    \itembf{Scalability} The objectives should not be severely degraded by an increasing number of users and reports.
    \itembf{Efficiency} The service should be efficient in terms of computation, memory, and communication overhead.
\end{itemize}

%------------------------------------------------------------------------------
\subsection{Security Objectives}
%------------------------------------------------------------------------------

\begin{itemize}
    \itembf{Secure communication} Reports and votes must be delivered confidential, authentic, and of integrity.
    \itembf{Resilient decision making} The service should be resilient against malicious reports and votes.
        Consequently, users must only issue one report about an event and vote once for each report.
        Thus, attackers must not be able to perform Sybil attacks. 
    \itembf{Accountability} Official authorities should be able to obtain the identity of a reporter or witness for the prosecution of crimes.
        However, restrictions must apply for access to this information in order to prevent abuse.
    \itembf{Availability} The verification service should provide resistance against \ac{DoS} attacks.
        This includes spamming of reports and votes.
\end{itemize}

%------------------------------------------------------------------------------
\subsection{Privacy Objectives}
%------------------------------------------------------------------------------

\begin{itemize}
    \itembf{Anonymity} Attackers must not learn about the identities of users issuing reports and votes.

    \itembf{Location privacy} Attackers must not determine the location of users.
        Otherwise, by following their movements, attackers might be able to infer their identities.

    \itembf{Co-location privacy} Attackers must not determine whether two users have been residing at the same location at the same time.
        Otherwise, attackers might infer a social connection between those users.

    \itembf{Absence privacy} Attackers must not learn about a user's absence from a location during a certain time.
        This information can be harmful if a user was not present at a location although he was supposed to be.
\end{itemize}

%------------------------------------------------------------------------------
\section{Verification Approach}
\label{sec:approach}
%------------------------------------------------------------------------------

In this article, we present a verification scheme, which we refer to as \acf{VUE}.
Our approach allows users to report events to one of potentially many \emph{verifiers} via their smartphones, i.e., \emph{\acp{UE}}.

In order to verify a report, the verifier issues confirmation requests to users that have been residing close to the event at the time the report has been submitted.
Delivering these confirmation requests in a privacy-preserving manner while supporting delay-tolerant communication and deferring of votes requires a \ac{STM} scheme \cite{wozniak2012geocast}.
It is necessary to rely on this concept as employing a \ac{geocast} scheme would require witnesses to stay close to the place of the event, which is an unrealistic assumption.
Therefore, building upon the approach in \cite{wozniak2012geocast}, we rely on \emph{\acp{RP}} to deliver confirmation requests in a privacy-preserving manner.
This \ac{RP}-based approach requires that users poll \acp{RP} in regular time intervals using a \emph{token}~$\tau$ containing a \emph{key}~$K$ that has been negotiated at some location and time in the past.
To allow for extensive anonymity guarantees these tokens are negotiated between nearby users in a cryptographically secure manner.
Hence, in certain time intervals, users initiate the negotiation of a \emph{group key}~$K$ with all users that are currently in communication range.
Users may also forward the negotiation requests over several hops to increase the number of users within a group and therefore the number of potential witnesses for some event in the future.
Tokens are considered valid up to some time after their reception, e.g., for 5\;minutes.
When issuing a report to the verifier, users include their currently valid tokens to allow the verifier to deposit a confirmation request at specific \acp{RP} so that potential witnesses of an event are able to retrieve it.

Finally, witnesses having obtained a request can issue their vote to the verifier, which is then able to decide whether a report is true based on the majority of votes.
In order to prevent Sybil attacks and to allow users to only issue one report or vote for each event, an \emph{\ac{IS}} is necessary that is able to authenticate the identity of users.
Therefore, in order to issue reports or votes, users have to obtain a \emph{voting ticket}~$\lambda$ from the \ac{IS} first.
This ticket contains a \emph{vote identifier}~$\upsilon$ that is unique for each user and report. 
Here, it is important that the \ac{IS} does not obtain the report itself in order to protect the privacy of users.
Therefore, he issues $\lambda$ for a given $h(M)$, where $h(x)$ is a cryptographic hash function and $M$ the message of the report.

\begin{figure}[!t]
\centering
\includegraphics{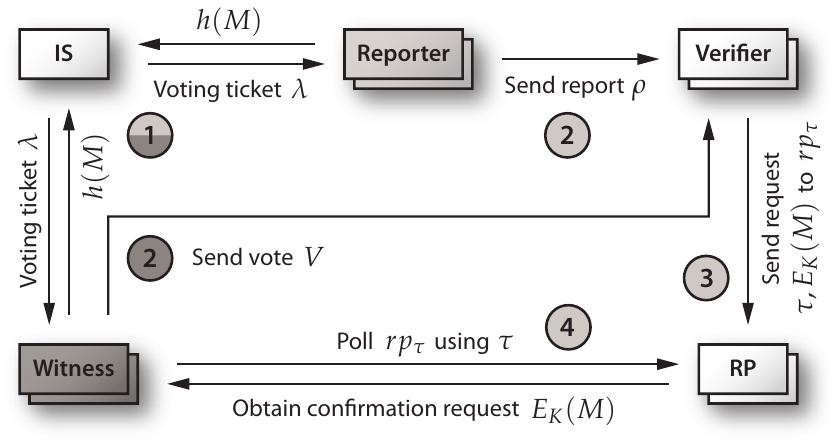}
\caption{Overview of \ac{VUE} approach. The process of issuing a report is shown in light gray, while issuing a vote is shown in dark gray.}
\label{fig:overview}
\end{figure}

We now provide a detailed overview of the four phases of the \ac{VUE} approach in the following sections (see Fig.~\ref{fig:overview}).

%------------------------------------------------------------------------------
\subsection{Token Negotiation}
%------------------------------------------------------------------------------

In the first phase, users initiate a secure protocol for a group key exchange in order to negotiate a common key $K$ (e.g. using \ac{GDH} \cite{steiner1996diffie}) in certain, randomly distributed time intervals within their $k$-hop neighborhood.
When the protocol is finished, users have negotiated a token ${\tau = h(K)}$ which is stored as a pair of $(\tau,K)$ and used to poll for confirmation requests later on.

%------------------------------------------------------------------------------
\subsection{Event Reporting}
%------------------------------------------------------------------------------

In order to report an event, a user has to interact with the \ac{IS} and verifier.
Hence, the user establishes a secure connection with the \ac{IS}, e.g., using the \ac{TLS} protocol, and sends $h(M)$ to the \ac{IS}.
Then, the \ac{IS} authenticates the identity of the user and responds with a voting ticket ${\lambda = (\upsilon,\left\{ h(M), \upsilon \right\}_{IS})}$, containing the unique vote identifier $\upsilon$.
Here, ${\upsilon = h(id, h(M), K_{IS})}$ with $id$ representing an identifier for the identity of the user, e.g. his \ac{IMSI}.
Furthermore, $\left\{ \right\}_{IS}$ is the public-key signature of \ac{IS} and $K_{IS}$ is a secret that is only known to the \ac{IS} and used to prevent the guessing of $\upsilon$ for a known identity and report message.

In order to prevent duplicate reports from different users, the reporting application should provide users with information about reports issued in their neighborhood, allowing them to recognize existing reports.
In this case, no additional report is sent, allowing users to proactively confirm this report so that if a confirmation request is received later on, the device can reply to the request without requiring further interaction from the user.

Then, the user establishes a secure connection with the verifier and sends his report ${\rho = ( \lambda, M, \alpha_1, \ldots, \alpha_l )}$, where ${M = (r, x, y, t, m)}$ contains the location~$x,y$, time~$t$, message description~$m$, and random number~$r$, which is used to prevent guessing of $h(M)$ by the \ac{IS}.
Furthermore, ${\alpha_i = (\tau_i, E_{K_i}(M))}$ represents the tokens $\tau$ and report messages $M$ which are symmetrically encrypted with the group key $K$ for all $t$ currently valid tokens.
Finally, the verifier computes $h(M)$ to verify the signature of the \ac{IS} and checks whether there is already a vote or report for the given vote identifier $\upsilon$.
If this is not the case, the report is accepted.

It should be noted here that it is possible to maintain multiple verifiers in order to provide resistance against attacks or to filter reports.
Therefore, police, fire, and ambulance department could each maintain their own verifiers.

%------------------------------------------------------------------------------
\subsection{Confirmation Request}
%------------------------------------------------------------------------------

In order to be able to decide whether a report is trustworthy, the verifier may send a confirmation request to specific \acp{RP}, where potential witnesses are able to retrieve them.
Therefore, for each $\alpha_i$ contained in the report, the verifier computes a \ac{RP} identifier $rp_{\tau_i} = h(\tau_i)$.
Like in \cite{wozniak2012geocast}, this identifier is used to obtain the name of the \ac{RP} where the request should be deposited.
By appending the number $rp_{\tau_i} \bmod N$ to a known prefix ($N$ is the number of \acp{RP}), the verifier can resolve the IP address of the \ac{RP}, e.g. by \ac{DNS}.
Finally, having established a secure connection, the verifier sends $(\tau_i, E_{K_i}(M))$ to the respective \ac{RP}, which stores $E_{K_i}(M)$ for lookup with $\tau_i$.

%------------------------------------------------------------------------------
\subsection{Witness Feedback}
%------------------------------------------------------------------------------

Witnesses poll \acp{RP} in regular time intervals to retrieve requests concerning their stored tokens $\tau$.
Here, the addresses of the \acp{RP} are derived as described in the previous section.
Once a user receives $E_{K}(M)$ for a token $\tau$, he decrypts it using the stored group key $K$ and decides about $M$.
Then, he establishes a secure connection to the \ac{IS} and obtains a voting ticket $\lambda$ as described above.
Finally, after having established a secure connection, he sends his vote ${V = (\lambda, \delta)}$ to the verifier, where ${\delta \in \left\{ \mbox{true}, \mbox{false}, \mbox{unsure}, \mbox{defer} \right\}}$ is his decision about the report.
Here, in order to support postponing of votes, if a device does not receive an input from the user within a certain time limit, it auto-replies with a \emph{defer}.
This allows the verifier to detect that a vote from a legitimate witness is still missing in order to postpone his decision making if a large number of votes is still missing.
If the verifier has received several votes providing a clear majority for the validity of a report, it is considered true.

%------------------------------------------------------------------------------
\section{Discussion}
\label{sec:discussion}
%------------------------------------------------------------------------------
\subsection{Security Aspects}
%------------------------------------------------------------------------------

Regarding the security objectives, we assume that potential attackers have one or more of the following goals: to obtain knowledge of the content of reports, its reporters and witnesses (see privacy discussion below), or to propagate misleading information in order to, e.g., impede relief operations or hide crimes.
To achieve these goals, attackers may observe the communication between entities, send reports, vote as a witness, or even compromise \acp{RP}.
However, attackers cannot compromise verifiers, the \ac{IS}, or parts of the cellular network infrastructure.
We consider this an appropriate assumption as it may be easier to control access to few verifiers or the \ac{IS} than protecting a large number of \acp{RP}, which may be required for scalability reasons.
With these abilities, we now discuss the given security objectives.

\subsubsection{Secure communication} 
Confidentiality, authentication, and integrity is provided by a protocol like \ac{TLS} that is employed between the entities.
Hence, by observing the communication or participating in the service adversaries cannot violate this objective.
It can also not be violated by compromising \acp{RP}, as those only store encrypted messages.

\subsubsection{Resilient decision making} 
By employing an \ac{IS}, attackers are not able to perform a Sybil attack and can only issue one report or vote per event.
Therefore, by participating in the service, they can only obtain a malicious majority, if the majority of votes is malicious.
While an adversary may try to issue false reports where he holds a malicious majority (i.e. by using non-existing tokens to exclude benign witnesses), this does not provide an advantage as long as benign users issue reports about the same event.
A more sophisticated attacker may also be able to compromise \acp{RP}.
In this case, while he may not manipulate votes directly, he can suppress confirmation requests to reduce the number of witnesses.
Nevertheless, if a report contains more than one token, requests are distributed to different \acp{RP} so that an adversary may only suppress a fraction of votes.
Finally, an adversary may try to manipulate decisions by identity theft, i.e., stealing votes.
Here, the only reasonable countermeasure is to implement a reputation scheme that allows to filter malicious or compromised \acp{UE}.
We plan to investigate such reputation-based filtering techniques in our future work.

\subsubsection{Accountability} 
While privacy of users is an important aspect, it still has to be possible to reveal the identity of a user for the prosecution of crimes.
This can be achieved by combining the knowledge of a verifier and the \ac{IS}, i.e., the vote identifier $\upsilon$ and $K_{IS}$, to infer the identity of a reporter or witness. 
However, due to pre-image resistance of $h(x)$, this still requires brute-force testing of all user identifiers $id$ and comparing it to ${\upsilon = h(id, ...)}$.
Hence, uncovering the identity of users is possible, but requires significant effort.

\subsubsection{Availability} 
Verifiers and the \ac{IS} can implement countermeasures against spamming by rejecting users who send reports or votes at too high rates.
Countermeasures against \ac{DoS} attacks may include techniques like, for example, client puzzles but are beyond the scope of this paper.

%------------------------------------------------------------------------------
\subsection{Privacy Aspects}
%------------------------------------------------------------------------------

In terms of the privacy objectives, we assume that potential attackers have one or more of the following goals: to infer the identity of users, their locations, co-location of users, or absence of users from a location.
We assume that attackers have the same abilities as described above.
Given these abilities, we discuss potential attacks against privacy.

%------------------------------------------------------------------------------
\subsubsection{Observation Attack}
%------------------------------------------------------------------------------
If an attacker observes the communication between entities, he is not able to violate the anonymity of users as he can only see an encrypted traffic flow.
Information about the identities belonging to the involved addresses requires additional knowledge from the cellular operator.
An adversary is also not able to violate the location, co-location, or absence privacy of users.
While he may observe which \acp{UE} poll which \acp{RP}, this does not provide an advantage since \acp{RP} are responsible for many locations at different times in an unpredictable manner due to the pre-image resistance of $h(x)$ and ${rp_{\tau} = h(\tau)}$ \cite{wozniak2012geocast}.
Furthermore, observing the communication with a verifier or the \ac{IS} does also not violate any objective as the attacker cannot read $M$ due to the encrypted communication.

%------------------------------------------------------------------------------
\subsubsection{Participation Attack}
%------------------------------------------------------------------------------
When having access to valid \acp{UE} adversaries may send reports and votes.
This corresponds to the knowledge of an attacker about specific $\tau$, $K$, and $M$.

\paragraph{Anonymity} 
As described above, user identities can only be violated if location, co-location, or absence privacy are violated as traffic is relayed only encrypted.

\paragraph{Location privacy} 
Observing users polling \acp{RP} does not violate the location privacy as \acp{RP} are responsible for many locations at different times.
Knowing $\tau$ and thus which \ac{RP} is used to deliver a confirmation request does therefore not violate this objective.
However, if there is only one report and the attacker knows the location contained in $M$, he may violate the location privacy as he is able to detect users voting for this report.
Still, such a temporal correlation may not be easy to detect with many reports and users reacting at different times. 
Furthermore, attackers can only obtain information about one location at a specific time, which is unlikely to be sufficient for inferring their identities.
Nevertheless, if reports are only issued rarely, users should still contact the IS and verifiers regularly to obfuscate temporal correlations.

\paragraph{Co-location privacy}
According to the location privacy, this objective may only be violated if there is just one report.

\paragraph{Absence privacy}
An attacker may not violate this objective, as he may only detect absence from a location if a user does not poll a certain \ac{RP}.
This is unlikely, as ${rp_\tau = h(\tau)}$ evenly distributes the responsibility of \acp{RP} for different times and locations.
Therefore, having received several tokens, a user is likely to poll every \ac{RP}. 

%------------------------------------------------------------------------------
\subsubsection{Compromising \acp{RP}}
%------------------------------------------------------------------------------
More sophisticated attackers may also compromise one or more \acp{RP}.
This corresponds to obtaining knowledge of tokens $\tau$ being polled by users.

\paragraph{Anonymity} 
As described above, anonymity can only be violated if location, co-location, or absence privacy is violated as attackers only obtain IP addresses of the users.

\paragraph{Location privacy}
Since the tokens $\tau$ that are stored on the \ac{RP} do not reveal any information about location or time (this requires knowledge of the group key $K$ exchanged among users in the area), an attacker has to participate in the service in order to violate this objective.
That is, he has to obtain $M$ and the corresponding $\tau$, as well as group key $K$. 
In this case, he can infer the \ac{RP} being polled by users having resided at that location at this time. 
If he is able to compromise this \ac{RP}, he can violate the location privacy of users having resided at the time and place contained in $M$.
Still, this is again not likely to be sufficient to infer the identity of users.
In order to track the movement of users, an attacker has to know several tokens $\tau$ received by a user which is only possible if he has been able to follow the user in the disaster area over some time.
Moreover, he has to be able to compromise several \acp{RP} to follow the movement.

\paragraph{Co-location privacy} 
An attacker may violate this objective as he can detect whether two users poll the same \ac{RP} using the same $\tau$. 
Nevertheless, this only provides knowledge of a potential social connection between two unknown users which may not be of much benefit. 
Assuming that an attacker wants to find out if two known users (given their IP addresses) have met, he has to be able to compromise a specific \ac{RP} responsible for the assumed place and time of the meeting.
In order to avoid this potential attack, users may want to use an anonymous proxy when polling \acp{RP} to hide the actual IP addresses.

\paragraph{Absence privacy}
By observing whether a user never polls a certain $\tau$ on a \ac{RP}, an attacker can violate the absence privacy of a user.
Nevertheless, this only provides an advantage if the user is known and the attacker is able to compromise a specific \ac{RP} for the location.
As discussed above, users may prevent this by using an anonymous proxy.

% Finally, it should be noted here, that the use of a \emph{token hierarchy} suggested in \cite{wozniak2012geocast} also contributes to prevent the violation of the privacy objectives, as in this case, an attacker has only a limited time to compromise a \ac{RP} before another \ac{RP} becomes responsible.

% %------------------------------------------------------------------------------
% \subsubsection{Privacy towards infrastructure}
% %------------------------------------------------------------------------------
% Apart from the attacks described in the previous sections, the question remains what the infrastructure might infer about users.
% With the \ac{IS} being aware of the identities of users, he might only learn about location, co-location, or absence of users.
% However, as the \ac{IS} is not able to guess $M$ due to random number $r$ in $M$, he is not able to infer any of these objectives.
% Verifiers only know about a few specific locations and times from issued reports and are therefore not able to fully track the movement of users.
% Therefore, according to accountability objective discussed above, only the combined knowledge of verifiers and \ac{IS} can violate the privacy of users.

%------------------------------------------------------------------------------
\section{Evaluation}
\label{sec:evaluation}
%------------------------------------------------------------------------------

\begin{table}[!t]
% \renewcommand{\arraystretch}{0.8}
% \setlength{\extrarowheight}{3pt}

\caption{Overview of simulation parameters}
\label{tbl:parameters}

% if using array.sty, it might be a good idea to tweak the value of
% \extrarowheight as needed to properly center the text within the cells
\footnotesize
\centering
% Some packages, such as MDW tools, offer better commands for making tables
% than the plain LaTeX2e tabular which is used here.
\begin{tabular}{|l|l|}
\hline
\bf Parameter & \bf Value \\
\hline
Simulated time & 120\;min (mobility warm-up 60\,min)\\
Field setup & 5\,$\times$\,5\;km$^2$, 2000 nodes, 100 events \\
Event radius & $\mathcal{U}$(25\;m, 250\;m) \\
Token negotiation interval & $\mathcal{U}$(15\;min, 30\;min) \\
Negotiation hop limit & 1, 2, 3, 4, 5, 6 \\
Token validity period & 5\;min (starting at time of reception) \\
Ratio of malicious nodes & 0\,..\,0.4 in steps of 0.05 \\
\hline
\bf Mobility Models & \acs{RWP}, \acs{RPGM}, \acs{NC} \\
Movement speed & $\mathcal{U}$(0.5\;m/s, 1.5\;m/s) \\
Group size (\acs{RPGM}, \acs{NC}) & $\mathcal{N}$($\mu$ = 4, $\sigma^2$ = 4) \\
% Group change prob. (\acs{RPGM}) & 0.1 \\
Max. pause duration & 60\;s (\acs{RWP}, \acs{RPGM}), 15\;min (\acs{NC}) \\
Max. group\,/\,roaming radius & 5\;m (\acs{RPGM}), 25\;m (\acs{NC}) \\
\hline
\bf Radio Model & IEEE 802.11 (2.4\;GHz, 54\;Mbit/s) \\
Transmit power & 17\;dBm (max. range $\approx$ 100\;m) \\
% Receiver sensitivity & \phantom{0}-65\;dBm \\
% Signal attenuation threshold & \phantom{0}-84\;dBm \\
% Thermal noise & -100\;dBm \\
Path loss model & log-distance, log-normal shadowing \\
Path loss coefficients & $n$ = 3.0, $\sigma$ = 9.5\;dB \\
Fast fading model & Jakes' Rayleigh fading \\
% Payload length & 1024\;bytes \\
\hline
\end{tabular}
\end{table}

In order to evaluate the performance of our \ac{VUE} approach, we implemented it in OMNeT++ \cite{varga2001omnet++} using the MiXiM framework \cite{wessel2009mixim}.
An overview of the simulation parameters is given in Table\;\ref{tbl:parameters}.
In our simulation, users move on a field according to a given mobility model while negotiating tokens within their $k$-hop neighborhood.
For replicability, we modeled events by circles with a given radius and users reporting an event when entering its area.
As we were interested in the number of witnesses that can be expected for a report when negotiating tokens over $k$ hops, we used three different mobility models: the well-known \acf{RWP}, as well as two group mobility models: \acf{RPGM} and \acf{NC} \cite{camp2002survey}.
We chose these models over existing mobility models for disasters since these models either do not consider the mobility of the affected population in a disaster or model it by applying existing models for group movement \cite{uddin2009post}.
Furthermore, in order to get an impression of the abilities of our approach in terms of detecting malicious users, we randomly set a certain fraction to be malicious.
At the end of the simulation, we collected the sets of witnesses for the reported events and calculated the ratio of benign majorities by counting the number of reports with more benign users and dividing it by the total number of reports.
We used this ratio as it corresponds to the correct confirmation of either true, or the rejection of false reports.
For witnesses, we assumed that malicious users are always able to vote for a malicious and against a benign report.
In contrast, benign witnesses only confirm a true or reject a false report, if they were actually in the event area.
Otherwise, they issue a vote with an unsure decision.

\begin{figure*}[!t]
\centering
\includegraphics{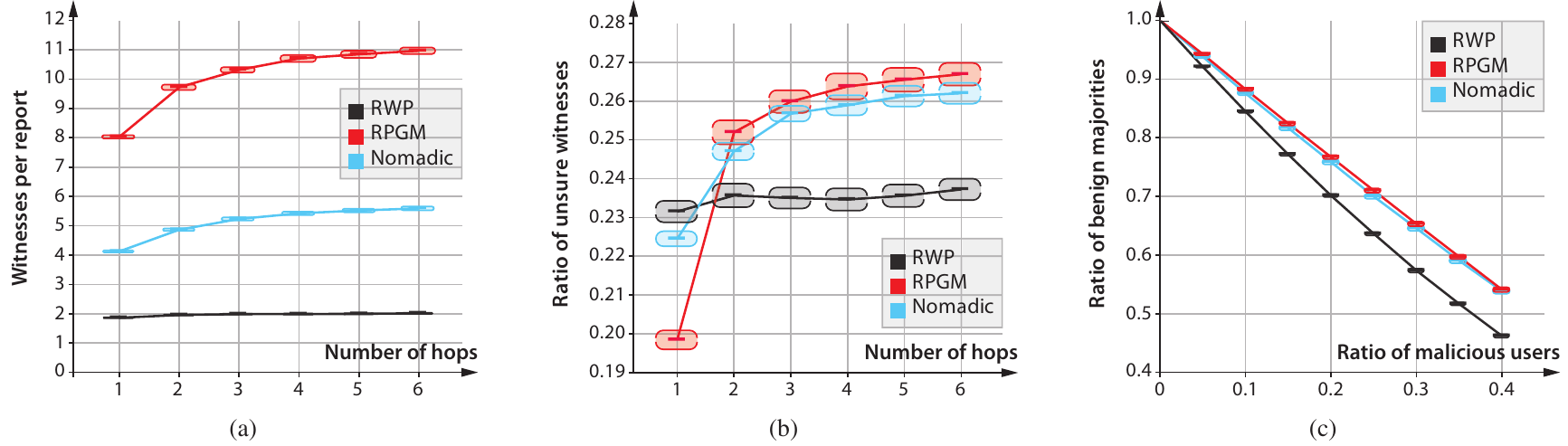}
\caption{Simulation results (average of 100 repetitions with 99\% confidence intervals)}
\label{fig:plots}
\end{figure*}

As expected, for an increasing number of $k$ hops, we can see that the number of witnesses increases as well (Fig.\;\ref{fig:plots}a).
Here, we can also see the impact of the different mobility models.
While for \ac{RWP}, were users just move randomly, the average number of witnesses is rather small with about 2 and only increases slightly, both group mobility models show a larger number of witnesses (between 4.2 and 5.5 for \ac{NC} and between 8 and 11 for \ac{RPGM}) and a stronger increase with increasing $k$.
This behavior can be explained with the group mobility models providing a larger number of witnesses through the movement in groups which provides a higher number of nodes that are in communication range.
Accordingly, we can see that the \ac{NC} model, where users move up to 25\,m away from the center of the group, the number of witnesses is smaller compared to the \ac{RPGM} model where users move closely to each other with only about 5\,m from the center of the group.
Since group mobility is more likely to appear after a disaster, we can see that our approach benefits from this with more witnesses per event.

Furthermore, according to our expectations, the ratio of unsure witnesses increases with the number of hops (Fig.\;\ref{fig:plots}b).
An interesting aspect here is the fact that for the group mobility models, at 1 hop, the ratio of unsure witnesses is smaller than for \ac{RWP}.
For more than 1 hop, the group mobility models suffer from the fact that users move in groups.
Here, it is more likely that witnesses using the same token have not been to the event area and are therefore unsure.
Hence, negotiating group keys over multiple hops does not seem to be a good strategy for disasters where users are likely to move in groups.

Finally, regarding the ratio of benign majorities for $k=1$ (Fig.\;\ref{fig:plots}c), we can see that the group mobility models are able to provide a higher ratio of benign majorities.
This behavior can be explained with the higher number of witnesses per report.
Nevertheless, for all mobility models, the approach suffers from benign users being unsure about an event.
Therefore, we can see that, e.g., for 10\% of malicious users, less than 90\% of all reports have a benign majority.

%------------------------------------------------------------------------------
\section{Conclusion}
\label{sec:conclusion}
%------------------------------------------------------------------------------

In this article, we proposed the concept of witness-based report verification.
We provided an extensive overview of objectives to be fulfilled by such a scheme and presented a first approach.
Our evaluation shows the benefit of group mobility, which results in a reasonable number of witnesses per event while relying on single-hop negotiation of tokens.

In our future work, we plan to investigate the impact of node densities and realistic reporting behavior that includes the aspect of witnesses voting at different times.
Finally, we aim to consider incorporating reputation schemes to filter malicious users and provide a comparison with existing verification schemes that are based on data mining techniques.


\section*{Acknowledgment}
%------------------------------------------------------------------------------
This work is supported by the German Research Foundation (DFG Graduiertenkolleg~1487, Selbstorganisierende Mobilkommunikationssysteme f\"ur Katastrophenszenarien).



%\bibliographystyle{IEEEtran}
% \end{thebibliography}

\bibliographystyle{IEEEtran}
\bibliography{IEEEabrv,references}

\end{document}